\documentclass[structabstract]{aa}  
\usepackage{graphicx}
\usepackage{txfonts}
\begin{document}
   \title{Study of star-forming galaxies in SDSS up to redshift 0.4\\
I. Metallicity evolution}

   \author{M.A. Lara-L\'opez
          \inst{1}
          \and
          J. Cepa\inst{1,2}
	  \and
          A. Bongiovanni\inst{1}
	  \and
	  A.M. P\'erez Garc\'{\i}a\inst{1}
	  \and
          H. Casta\~neda\inst{1,3}
	  \and
          M. Fern\'andez Lorenzo\inst{1}
	  \and
          M. Povi\'c\inst{1}
	  \and
          M. S\'anchez-Portal\inst{4}
         }

   \institute{Instituto de Astrof\'{\i}sica de Canarias, 38200 La Laguna, Spain
              \email{mall@iac.es}
         \and
             Departamento de Astrof\'{\i}sica, Universidad de la Laguna, Spain
          \and
            Departamento de F\'{\i}sica, Escuela Superior de F\'{\i}sica y Matem\'atica, IPN, M\'exico D.F., M\'exico
          \and
             Herschel Science Center, INSA/ESAC, Madrid, Spain
             }

   \date{Received; accepted}

\abstract  
{The chemical composition of the gas in galaxies versus cosmic time provides a very important tool for understanding galaxy evolution. Although there are many studies at high redshift, they are rather scarce at lower redshifts. However, low redshift studies can provide important clues about the evolution of galaxies, furnishing the required link between local and high redshift universe. In this work we focus on the metallicity of the gas of star--forming galaxies at low redshift, looking for signs of chemical evolution.} 
{To analyze the metallicity contents  star--forming galaxies of similar luminosities and masses at different redshifts. With this purpose, we present a study of the metallicity of relatively massive (log(M$_{star}$/M$_{\odot}$) $\gtrsim$ 10.5) star forming galaxies from SDSS--DR5 (Sloan Digital Sky Survey--Data Release 5), using different redshift intervals from 0.04 to 0.4.}
{We used data processed with the STARLIGHT spectral synthesis code, correcting the fluxes for dust extinction, estimating metallicities using the $R_{23}$ method, and segregating the samples with respect to the value of the [{N\,\textsc{ii}}] $\lambda$6583/[{O\,\textsc{ii}}] $\lambda$3727 line ratio in order to break the $R_{23}$ degeneracy selecting the upper branch. We analyze the luminosity and mass-metallicity relations, and the effect of the Sloan fiber diameter looking for possible biases.}
{By dividing our redshift samples in intervals of similar magnitude and comparing them, significant signs of metallicity evolution are found. Metallicity correlates inversely with redshift: from redshift 0 to 0.4 a decrement of $\sim$0.1 dex in 12+log(O/H) is found.}
{}
 \keywords{galaxies: abundances --
                galaxies: evolution --
                galaxies: starburst
               }

\maketitle { }
%

\section{Introduction}

Determination of the chemical composition of the gas and stars in galaxies versus cosmic time provides a very important tool for understanding galaxy evolution, due to its important impact on fields such as stellar evolution and nucleosynthesis, gas enrichment processes, and the primary or secondary nature of the different chemical species. Historically, the main observational evidence suggestive of chemical evolution of galaxies, is provided by the observation of different chemical compositions of stars of different ages of the Milky Way and its environment (see the reviews by Audouze $\&$ Tinsley 1976, Wheelet et al. 1989, Wilson $\&$ Matteucci 1992, McWilliam 1997).

The study of the evolution of the metal enrichment in galaxies is based mainly in two methods. One is based on the detection of absorption lines in QSO spectra produced by the neutral interstellar medium (ISM) of galaxies in the line--of--sight of the QSO (Prochaska et al. 2003), while the other uses emission lines of the warm ISM ({H\,\textsc{ii}} regions) detected in the integrated galaxy spectra.

Optical emission lines in galaxies have been widely used to estimate abundances in extragalactic {H\,\textsc{ii}} regions (e.g. Aller 1942; Searle 1971; Pagel 1986; Shields 1990, among others). Among the different methods developed  to estimate metallicities, we can distinguish between theoretical models, empirical calibrations, or a combination of both (for a review see, e.g., Kewley $\&$ Dopita 2002; Kewley $\&$ Ellison 2008). The direct method to estimate metallicities in galaxies is known as the $``$$T_e$ method$"$ (Pagel et al. 1992; Skillman $\&$ Kennicutt 1993), which consists on measuring the ratio of the [{O\,\textsc{iii}}] $\lambda$4363 auroral line to a lower excitation line such as [{O\,\textsc{iii}}] $\lambda$5007. Assuming a classical {H\,\textsc{ii}} region model, this ratio provides an estimate of the electron temperature of the gas, which is then converted into metallicity (Osterbrock 1987). However, [{O\,\textsc{iii}}] $\lambda$4363 is too weak to be easily observed, not only in metal rich, but even in metal poor galaxies, Z $<$ 0.5 $Z_ \odot$ (log(O/H) + 12 $<$ 8.6), and according to Kobulnicky et al. (1999), for low--metallicity galaxies, the [{O\,\textsc{iii}}] $\lambda$4363 diagnostic systematically underestimates the global oxygen abundance. Also, the same authors show that for massive, metal--rich galaxies, empirical calibrations using strong emission--line ratios can be reliable indicators of the overall oxygen abundance in {H\,\textsc{ii}} regions.

For these reasons, theoretical metallicity calibrations of strong--line ratios using photoionization models are used instead  for determining abundances of high metallicity star--forming galaxies, such as: [{N\,\textsc{ii}}] $\lambda$6584/[{O\,\textsc{ii}}] $\lambda$3727 (Kewley $\&$ Dopita 2002) and the $R_{23}$ ratio, introduced by Pagel et al. (1979). The first one provides an excellent abundance diagnostic for Z $>$ 0.5 $Z_ \odot$ (log(O/H) + 12 $\gtrsim$ 8.6), because $\rm{N}^{+}$ and $\rm{O}^{+}$ have similar ionization potentials, and this ratio is almost independent of ionization parameter. However it cannot be used at lower abundances (Z $<$ 0.5 $Z_ \odot$), where the metallicity dependence of the [{N\,\textsc{ii}}] $\lambda$6584/[{O\,\textsc{ii}}] $\lambda$3727 ratio is lost because nitrogen (like oxygen) is predominantly a primary nucleosynthesis element in this metallicity range (Kewley $\&$ Dopita 2002).
The $R_{23}$ method, is a widely used and well calibrated method (see for example Alloin et al. 1979; Edmunds $\&$ Pagel 1984;  McCall et al. 1985; Dopita $\&$ Evans 1986; McGaugh 1991; Zaritsky et al. 1994; Kewley $\&$ Dopita 2002; Kobulnicky $\&$ Kewley 2004; Tremonti et al. 2004, hereafter T04; Liang et al. 2006, hereafter L06). However, it has the disadvantages of being double--valued as a function of 12$+$log(O/H), and that depends on the ionization parameter, particularly for Z $<$ 0.5 $Z_ \odot$, being less sensitive to metallicity in this range.

Alternatively, when the direct method can not be used, empirical calibrations can be obtained by fitting the relationship between direct $T_e$ metallicities and strong--line ratios as well. Typical empirical calibrations are: the $R_{23}$ ratio (Pilyugin 2001; Pilyugin $\&$ Thuan 2005; Liang et al. 2007), from which Pilyugin (2001) derived an empirical calibration based on $T_e$ metallicities for a sample of {H\,\textsc{ii}} regions, the [{N\,\textsc{ii}}] $\lambda$6583/ {H$\alpha$} ratio (Pettini $\&$ Pagel 2004, hereafter PP04), and the ([{O\,\textsc{iii}}] $\lambda$5007/{H$\beta$})/([{N\,\textsc{ii}}] $\lambda$6583/{H$\alpha$}) ratio (O3N2), (PP04). Although the latest method is of little use when O3N2 $\gtrsim$ 2, at lower values the relation is relatively tight and linear (PP04).

Finally, as an example of a combined calibration, we have the N2 = [{N\,\textsc{ii}}] $\lambda$6583/ {H$\alpha$} method (Denicol\'o et al. 2002), which follows a linear relation with log(O/H)  that holds for approximately from 1/50th to twice the Solar value. This method is based on a fit to the relationship between the $T_e$ metallicities and the [{N\,\textsc{ii}}] $\lambda$6583/{H$\alpha$ ratio, from which some have metallicities derived using the $T_e$ method, and the remaining metallicities were estimated using either the theoretical $R_{23}$ or an empirical method.

Nevertheless, comparisons among the metallicities estimated using different theoretical and empirical methods reveal large discrepancies (e.g., Pilyugin 2001, Bresolin et al. 2004, Garnett et al. 2004), with theoretical calibrations favouring higher metallicity values than those obtained using electron temperature estimations.

In the field of metallicity evolution versus cosmic time, there exist many studies, both theoretical and observational. Among the models we have, for example, that of Buat et al. (2008) and Kobayashi et al. (2007), who derived models of metallicity as a function of $z$, which show a progressive increase in metallicity with time, even at low redshift. Savaglio et al. (2005), developed an empirical model of metallicity evolution based on observations, in which the metallicity  at z  $<$ 1 is an interpolation of that at higher redshifts. Also, Brooks et al. (2007) and Finlator $\&$ Dav\'e (2008), among others, derived cosmological models of the mass-metallicity relation.  The cosmic metal enrichment  is attributed to a higher past volume--averaged star formation rate (see for example Madau et al. 1996; Lilli et al. 1996; Flores et al. 1999).

In addition, the metallicity and masses of galaxies are strongly correlated, with massive galaxies showing higher metallicities than less massive galaxies. This mass-metallicity ($M-Z$) relation has been intensively studied (Skillman et al. 1989; Brodie $\&$ Huchra 1991; Zaritsky et al. 1994; Richer $\&$ McCall 1995; Garnett et al. 1997; Pilyugin $\&$ Ferrini 2000, among others), and it is well established in the local universe (z $\sim$ 0.1) by the work of T04 using SDSS data. Regarding to this evolution of the mass--metallicity relation of star--forming galaxies at high redshift, Erb et al. (2006) found that star--forming galaxies at redshift  $\sim$ 2 have 0.3 dex fainter metallicities. Similarly, Maiolino et al. (2008) found evolution at  z $\sim$ 3.5, which appear to be much stronger than the one observed at lower redshifts, suggesting that this redshift corresponds to an epoch of major activity in terms of star formation and metal enrichment. At intermediate redshifts (1 $<$ z $<$ 2),  there are several important studies of the evolution of the chemical composition of the gas, such as the ones by Maier et al. (2006), P\'erez-Montero et al. (2009), and Liu et al. (2008), the last one found that the zero point of the $M-Z$ relation evolves with redshift, in the sense that galaxies at fixed stellar mass become more metal--rich at lower redshift.

Among the studies at z $<$ 1, usually based on small samples,  Savaglio et al. (2005) have investigated the mass--metallicity relations using galaxies at 0.4 $<$ z $<$ 1, finding that metallicity is lower at higher redshift, for the same stellar mass, by $\sim$ 0.15 dex. Also, Maier at al. (2005), from a sample of 30 galaxies with 0.47 $<$ z $<$ 0.92, found that one--third have  metallicities lower than those of local galaxies with similar luminosities and star formation rates. Consistently, Hammer et al. (2005) and Liang et al. (2006) found that at z $\sim$ 0.7, emission line galaxies were poorer in metals than present--day spirals by 0.3 dex. However, Kobulnicky $\&$ Kewley (2004) report a smaller variation of 0.14 dex for 0 $<$ z $<$ 1. This difference could be attributed to the fact that the last authors used equivalent widths and standard underlying stellar absorption, rather than high quality calibrated spectra for measuring the Balmer absorption. Lilly et al. (2003), from a sample of 66 star forming galaxies with 0.47 $<$ z $<$ 0.92, found a smaller variation in metallicity of $\sim$ 0.08 dex compared with the metallicity observed locally, showing only modest evolutionary effects. On the contrary, Carollo $\&$ Lilly (2001), from emission--line ratios of 15 galaxies in a range of 0.5 $<$ z $<$ 1, found that their metallicities appear to be remarkably similar to those of local galaxies selected with the same criteria. A similar result, consistent with no significant evolution, was found for the luminosity--metallicity relation by Lamareille et al. (2006), comparing 131 intermediate redshift star--forming galaxies (0.2 $<$ z $<$ 1, split in 0.2 redshift bins). However, a recent study of Lamareille et al. (2009), focussed on the evolution of the $M-Z$ relation up to z $\sim$ 0.9, suggesting that the $M-Z$ relation is flatter at higher redshifts. At z $\sim$ 0.77, galaxies of $10^{9.4}$ solar masses have $-0.18$ dex lower metallicities than galaxies of similar masses in the local universe, while galaxies of $10^{10.2}$ solar masses have $-0.28$ dex lower metallicities.

These discrepancies point to a need for studying the lower redshift galaxy samples, to ascertain whether or not at such low redshifts (i.e. an age of 8.4 Gyr for z $\sim$ 0.5, using a concordance $\Lambda$-CDM  cosmology, H$_{0}$=70,  $\Omega_m=0.3$ and  $\Omega_\Lambda=0.7$; Spergel et al. 2003) there exist evidences for metallicity evolution, as also to serve as calibrator of higher redshift studies. However, to be able to compare different redshift samples, it is advisable to use the same method for estimating metallicities, since, as explained above, theoretical and empirical calibrations generate discrepancies in the metallicity estimates depending on the used method.

The SDSS database provides an excellent opportunity for extending these studies down to z $\sim$0.4, in order to explore a possible evolution of metallicity at low--redshift, but using large samples, thus deriving more statistically significant results. In this paper, we extend the study presented in our previous article (Lara-L\'opez et al. 2009, hereafter L09), from 207 to more than 12000 galaxies, spanning more luminosity intervals in  redshift bins of 0.1 from $\sim$0  to 0.4, and adding analyses of the mass and luminosity-metallicity relations, as well as about the origin of nitrogen in our galaxies.

This paper is structured as follows: in Sect. 2 we give a detailed description of the used data, in Sect. 3 we describe the metallicity estimates, the [{N\,\textsc{ii}}] $\lambda$6583/[{O\,\textsc{ii}}] $\lambda$3727 diagram and its metallicity distribution, in Sect 4 we investigate the origin of nitrogen of our galaxies, whereas in Sect. 5 we discuss our results taking into account the possible biases of our samples. Finally, conclusions are given in Sect. 6.


\section{Sample selection}


We analyzed the properties of a selected sample of emission lines galaxies from SDSS--DR5 (Adelman--McCarthy et al. 2007). Targets were observed using a 2.5 m telescope located at Apache Point Observatory (Gunn et al. 2006). The SDSS spectra were obtained through 3 arcsec diameter fibres, covering a wavelength range of 3800-9200 {\AA}, and with a mean spectral resolution $\lambda$/$\Delta\lambda$ $\sim$ 1800. The SDSS--DR5 spectroscopy database contains spectra for $\sim$ $\rm{10}^6$ objects over $\sim$ 5700 $\rm{deg}^2$. Further technical details can be found in Stoughton et al. (2002).

We used the SDSS--DR5 spectra from STARLIGHT Database\footnote{http://www.starlight.ufsc.br}, which were processed through the STARLIGHT spectral synthesis code, developed by Cid Fernandes and colleagues (Cid Fernandes et al. 2005, 2007; Mateus et al. 2006; Asari et al. 2007). From them, we obtained the emission lines fluxes measurements of our samples from the contiuum subtracted spectra. For each emission line, STARLIGHT code returns the rest frame flux and its associated equivalent width, linewidth, velocity displacement relative to the rest--frame wavelength and the S/N of the fit. In the case of Balmer lines, STARLIGHT code corrects for underlying stellar absorption using synthetic spectra obtained by fitting an observed spectrum  with a combination of 150  simple stellar populations (SSPs) from the evolutionary synthesis models of Bruzual $\&$ Charlot (2003),  computed using a Chabrier (2003) initial mass function, $``$Padova 1994$"$ evolutionary tracks (Alongi et al. 1993; Bressan et al. 1993; Fagotto et al. 1994a, b; Girardi et al. 1996), and STELIB library (Le Borgne et al. 2003). The 150 base elements span 25 ages between 1 Myr and 18 Gyr, and six metallicities, from $Z=$0.005 to 2.5 $Z_{\odot}$; for more details see Mateus et al. (2006).

Our objective is to study the chemical evolution of emission--line galaxies by comparing galaxies at different redshifts in small and equal ranges of luminosities. To this aim, our sample is divided in redshift intervals of 0.1 from z $\sim$ 0 to 0.4. The purpose for selecting small ranges in luminosity is to alleviate the problem of the magnitude completeness, since the Sloan sample is only complete in the magnitude range $14.5 < m_r < 17.7$ (e.g., Asari et al. 2007), and then becomes incomplete at redshift above z $>$ 0.1 (e.g., Kewley et al. 2006). However, this procedure limits the study to the more luminous galaxies, since they are the ones detected at any redshift interval.

With this aim, our initial sample was divided in the following redshift intervals: $z_0$=(0.04$-$0.1), $z_1$=(0.1$-$0.2), $z_2$=(0.2$-$0.3) and $z_3$=(0.3$-$0.4). To ensure covering $>$ 20$\%$ of the light, we selected galaxies for the $z_0$ sub--sample with $z >$ 0.04 , as recomended by Kewley et al. (2005). This preliminary selection give us 197967 galaxies for $z_0$, 226012 for $z_1$, 42205 for $z_2$, and 38305 for the $z_3$ sub--samples. Absolute magnitudes were both K  and Galactic extinction corrected, by using the code provided by Blanton et al. (2003), and the maps of Schlegel et al. (1998), respectively, as provided by the STARLIGHT team.

In order to determine the luminosity--intervals at each redshift, we proceed to estimate the completeness of the absolute Petrosian r magnitude for each redshift--interval. We obtained our luminosity--complete samples selecting all galaxies brighter than the median for the larger redshift interval to be compared with, as shown in Fig. 1. According to that, we proceed constructing three luminosity--samples (hereafter l--samples): the l--sample $a$ was constructed by taking the interval of the  luminosity--completeness of $z_1$ ($-23.8 <$ M$_r <-21.7$), this allows comparing redshift--intervals $z_0$, and $z_1$; in l--sample $b$ we take the luminosity--completeness of $z_2$ ($-24.8 <$ M$_r <-22.9$), for comparing redshift--intervals $z_2$, and $z_1$; and in l--sample $c$ we take the luminosity--completeness of $z_3$ ($-24.8 <$ M$_r <-23.1$), to compare redshift--intervals $z_3$, $z_2$, and $z_1$, as shown in Table 1 and Fig. 1. As resulting from this figure, it was not possible to introduce the $z_0$ redshift--sample for comparison with l-samples $b$ and $c$, due to the small number of galaxies at their luminosity completeness.

\begin{table}[h]
\begin{center}
\begin{tabular}{cccc}
\hline
\hline
L-sample&redshift&$\rm{M}_r$&redshift-samples\\\
&completeness&& compared\\\hline
$a$&$z_1$&($-$23.8,$-$21.7)&$z_1$, $z_0$\\\
$b$&$z_2$&($-$24.8,$-$22.9)&$z_2$, $z_1$\\\
$c$&$z_3$&($-$24.8,$-$23.1)&$z_3$, $z_2$, $z_1$\\\hline

\end{tabular}
\normalsize
\end{center}
\caption{Luminosities--samples studied with their respective luminosity interval in absolute Petrosian r-magnitude, and the redshift--samples that can be compared.}
\end{table}

 \begin{figure}
\centering
\includegraphics[scale=0.60]{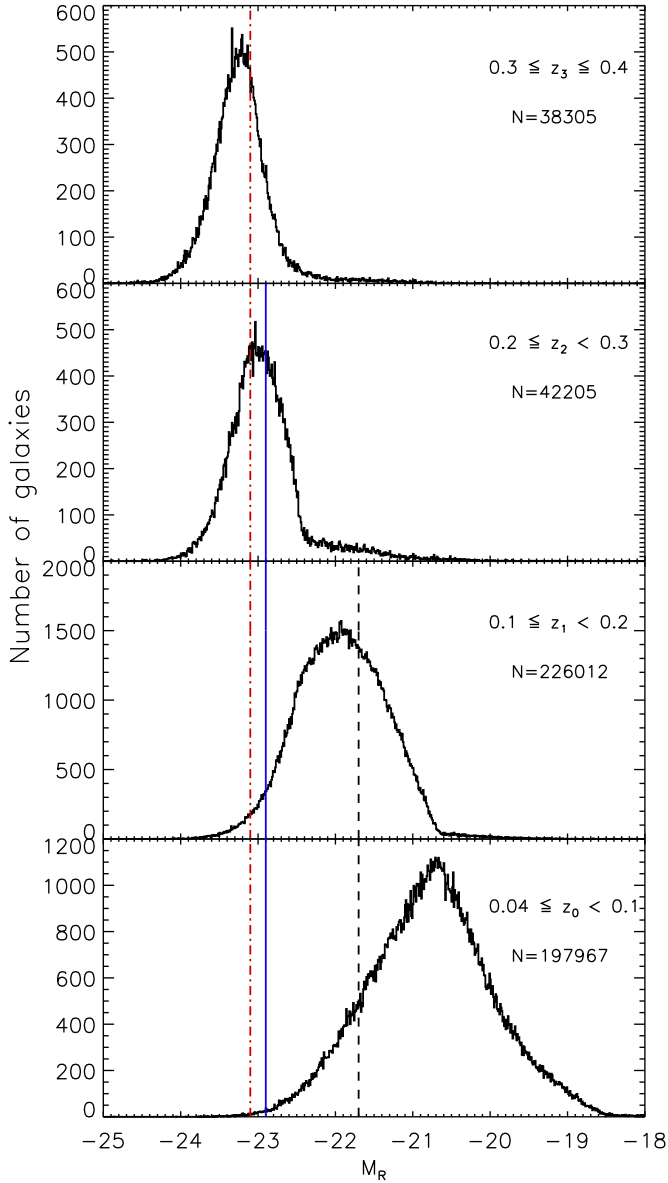}
\caption{Absolute Petrosian r--magnitude histograms. From top to bottom, histograms for the $z_3$, $z_2$, $z_1$, and $z_0$ sub--samples, respectively. Dot--dashed line represents the completeness for the $z_3$ sample, solid line shows the completeness for the $z_2$ sample, and dashed lines the completeness for the $z_1$ sample. All of them extended along the l-samples $a$, $b$ and $c$, where N is the number of galaxies, respectively.}
 \end{figure}

The resulting samples contain, for the l-sample $b$:  2352 galaxies for $z_2$ and  1386 for $z_1$; for l-sample $a$: 37777 for $z_1$ and  9288 for $z_0$. The l-sample $c$, was already studied in in L09.

From these samples we only consider galaxies whose spectra show in emission the {H$\alpha$}, {H$\beta$}, [{N\,\textsc{ii}}] $\lambda$6583, [{O\,\textsc{ii}}] $\lambda$3727, [{O\,\textsc{iii}}] $\lambda$4959 and [{O\,\textsc{iii}}] $\lambda$5007 lines, with {H$\alpha$}, {H$\beta$}, [{N\,\textsc{ii}}] $\lambda$6583 and  [{O\,\textsc{ii}}] $\lambda$3727 signal-to-noise ratios higher than 3$\sigma$. 

Finally, we selected star--forming galaxies following the criteria given by Kauffmann et al. (2003) in the Baldwin et al. (1981) diagram:  log[{O\,\textsc{iii}}] $\lambda$5007/{H$\beta$} $\le$ 0.61/$\{$log([{N\,\textsc{ii}}] $\lambda$6583/{H$\alpha$})-0.05$\}$ + 1.3,  used for example by Veilleux $\&$ Osterbrock (1987), Kewley et al. (2001, 2006), Stasi\'nska et al. (2006), among others. After all these selections, the number of galaxies for each sample is, for the l-sample $b$: 335 galaxies for $z_2$, and 148  for $z_1$, and for the l-sample $a$: 10477 for $z_1$, and 1577 for $z_0$.

\subsection{Dust extinction}

Since Balmer lines are already corrected for underlying stellar absorption by the STARLIGHT code,  it only remains correcting  for dust extinction. Our extinction correction was derived using the Balmer decrements in order to obtain the reddening coefficient C({H$\beta$}). Assuming case--B recombination with a density of 100 cm$^{-3}$ and a temperature of 10$^{4}$ K, the predicted ratio (unaffected by reddening or absorption) of {H$\alpha$}/{H$\beta$} is 2.86 (Osterbrock 1989), and the coefficient is given by:
 
 \begin{center}
C(H$\beta$)=${\frac{1}{f(\lambda)}}$log$\left[{\frac{I({\rm{H}\alpha})}{I({\rm{H}\beta})}}/{\frac{F({\rm{H}\alpha})}{F({\rm{H}\beta})}}\right]$,
\end{center} where F($\lambda$) and I($\lambda$) are the observed and the theoretical fluxes, respectively,  and f($\lambda$) is the reddening curve normalized to H$\beta$ using the Cardelli et al (1989) law, with $R_v$ = $A_v$/E(B -- V)=3.1

Once we have obtained the reddening coefficient for each galaxy of our samples, we proceed to estimate the corrected fluxes using  $F_{corr}$($\lambda$)= $F_{obs}(\lambda)10^{0.4 A_\lambda}$, with,

  \[
     \begin{array}{lp{0.8\linewidth}}
A_{\rm{H}\beta}=2.5{\;}C(H\beta)\\
A_{[{\rm{N}\,\textsc{ii}}] \lambda6583}=1.747{\;} C(H\beta)\\
A_{\rm{H}\alpha}=1.758{\;} C(H\beta)\\
A_{[{\rm{O}\,\textsc{iii}}] \lambda5007}=2.403{\;} C(H\beta)\\
A_{[{\rm{O}\,\textsc{iii}}] \lambda4959}=2.433{\;} C(H\beta)\\
A_{[{\rm{O}\,\textsc{ii}}] \lambda3727}=3.303{\;} C(H\beta),\\

         \end{array}
  \] as calculated from a prescription given by Cardelli et al. (1989)

\section{Metallicity estimates and evolution} 

We estimate metallicities using the $R_{23}$ relation, introduced by Pagel et al. (1979), 

 \begin{center}
$R_{23}$=([{O\,\textsc{ii}}] $\lambda$3727+[{O\,\textsc{iii}}] $\lambda\lambda$4959, 5007)/H$\beta$, 
\end{center} adopting the calibration given by Tremonti et al. (2004), 
\begin{equation}
12+\rm{log(O/H)} = 9.185 - 0.313x - 0.264x^2 - 0.321x^3, 
\end{equation}where $x$ = log $R_{23}$. 

However, this calibration is valid only for the upper branch of the double--valued $R_{23}$ abundance relation, and additional line ratios, such as [{N\,\textsc{ii}}] $\lambda$6583/[{O\,\textsc{ii}}]$\lambda$3727, are required to break this degeneracy. Since the upper and lower branches of the $R_{23}$ calibration bifurcate at log([{N\,\textsc{ii}}]/[{O\,\textsc{ii}}]) $\sim$ --1.2 for the SDSS galaxies (Kewley $\&$ Ellison 2008), which corresponds to a metallicity of 12+log(O/H) $\sim$ 8.4, we further select galaxies having  12$+$log(O/H) $>$ 8.4 and log([{N\,\textsc{ii}}]/[{O\,\textsc{ii}}]) $>$ --1.2, corresponding to the upper $R_{23}$ branch. Applying this final discrimination, we end for l-sample $b$ with 331 galaxies for $z_2$ and 146 for $z_1$,
 and for l-sample $a$ with 10434 galaxies for $z_1$ and 1576 for $z_0$. These are the samples that will be analyzed in this paper.

From these data, we derived the abundance--sensitive diagnostic diagram [{N\,\textsc{ii}}] $\lambda$6583/[{O\,\textsc{ii}}] $\lambda$3727 vs. 12+log(O/H), represented in Fig. 2. This diagram has also been used, for example, by Kewley $\&$ Dopita (2002), Nagao et al. (2006), and L06, among other metallicity--sensitive emission--line ratios, like log([{N\,\textsc{ii}}] $\lambda$6583/H$\alpha$), log([{O\,\textsc{iii}}] $\lambda$5007/H$\beta$)/[{N\,\textsc{ii}}] $\lambda$6583/H$\alpha$), and log([{O\,\textsc{iii}}] $\lambda$4959, 5007/H$\beta$). We selected this specific diagram due to its low scatter and to the physical additional information that provides. The advantages of using [{N\,\textsc{ii}}] $\lambda$6583 and [{O\,\textsc{ii}}] $\lambda$3727 lines, is that they are not affected by underlying stellar population absorption, and because this ratio is almost independent of the ionization parameter, since $\rm{N}^{+}$ and ${\rm{O}}^{+}$ have similar ionization potentials. From all  the diagnostic diagrams cited before, this one presents the lowest scatter, and since both axis are sensible to metallicity, possible signs of evolution could be easily identified in it.

\begin{figure}[ht!]
\includegraphics[scale=0.6]{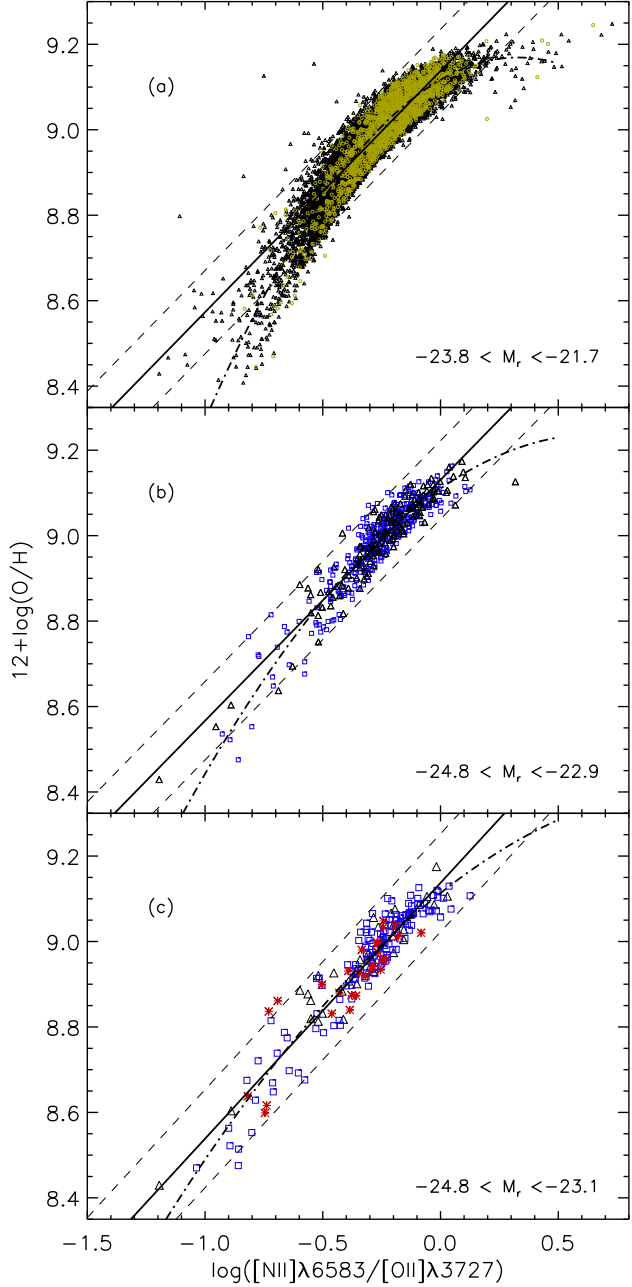}

\caption{Calibration relation between 12+log(O/H) and log([{N\,\textsc{ii}}] $\lambda$6583/[{O\,\textsc{ii}}] $\lambda$3727). From top to bottom, l-sample $a$, $b$, and $c$. Circles, triangles, squares and asterisks, represent galaxies for the redshift--sample $z_0$, $z_1$, $z_2$, and $z_3$, respectively. The solid line represents the best fit of the data using a linear fit, with the 2$\sigma$ dispersion indicated by the short--dashed lines, while the dot--dash line shows the order-3 polynomial fit.}
         \label{FigVibStab}
   \end{figure}

\begin{figure}[ht!]
\includegraphics[scale=0.6]{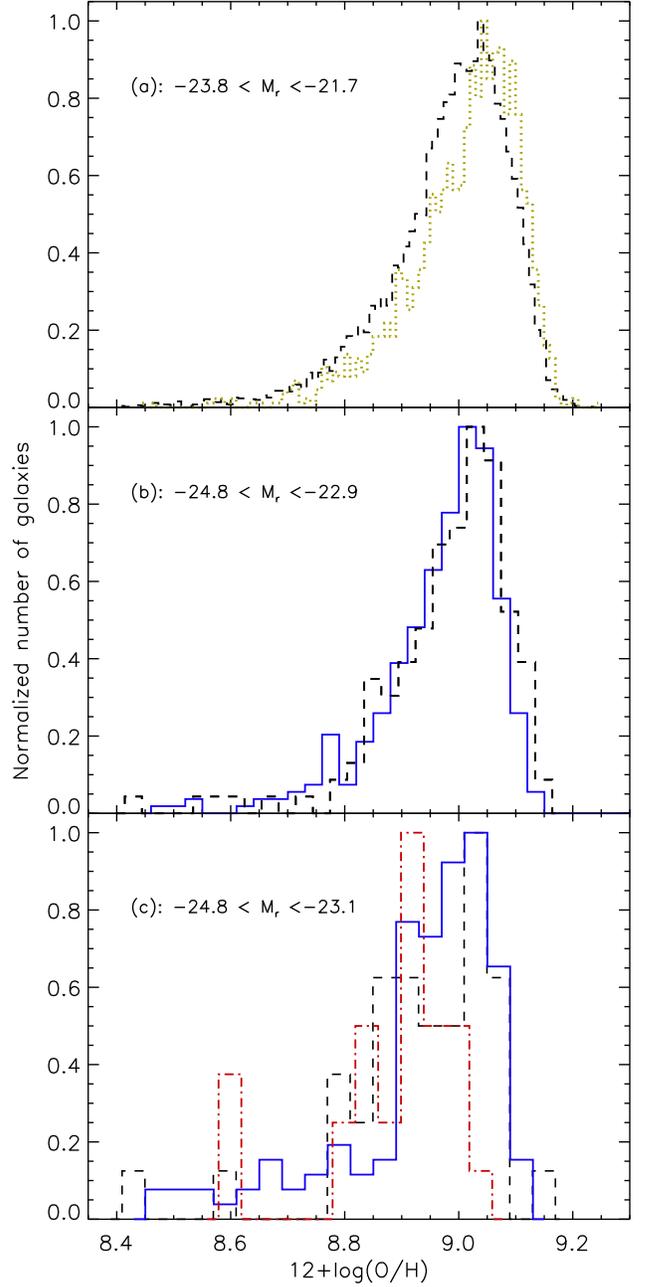}
\caption{Normalized metallicities histograms for our three l-samples. Dotted line represents the redshift--sample  interval $z_0$, dashed line the redshift--sample $z_1$, solid line the redshift--sample $z_2$ and dot-dash line the redshift--sample $z_3$.}
         \label{FigVibStab}
   \end{figure}

For every l-sample of Fig.2, we fit both a linear and an order three polynomial, obtaining the coefficients shown in Table 2.

\begin{table}[h]
\begin{center}
\begin{tabular}{ccccccc}   
\hline
\hline
{}&\multicolumn{2}{c} {\bf Linear Fits}&{}&\multicolumn{3}{c}{\bf Polynomial Fits}\\\cline{5-7}
\hline
l-sample&$b_0$&$b_1$&$\sigma$&$a_0$&$a_1$&$a_2$\\\hline
$a$&9.139&0.575&0.097&9.117&0.329& -0.447\\\
$b$&9.130&0.564&0.091&9.112& 0.408&-0.226\\\
$c$&9.137&0.599&0.114&9.114&0.438& -0.183\\\hline
\noalign{\smallskip}
\end{tabular}
\normalsize
\rm
\end{center}
\caption{Coefficients for the l-samples. For linear fits we assume $y=b_0+b_1x$, and for polynomial fits $y=a_0+a_1x+a_2x^2$, with $y$ = 12+log(O/H) and $x$ = log([{N\,\textsc{ii}}] $\lambda$6583/[{O\,\textsc{ii}}] $\lambda$3727).}
\vspace{0.1cm}
\end{table}

To interpret our results, it is important to note that we are working with the integrated spectra, and it is well known that the metallicity decreases with the distance from the galaxy center (Garnett et al. 1997). As shown by Kewley et al. (2005), and L06 using a sample of Jansen et al. (2000), data points from nuclear spectra follow the SDSS galaxies nuclear spectra very well, but data points from the integrated spectra show lower 12+log(O/H). Nuclear metallicities exceed metallicities derived from integrated spectra by $\sim$0.13 dex on average.

As shown in L09, for the l-sample $c$ there is a clear decrement in the $z_3$ redshift-sample of $\sim$0.1 dex in 12+log(O/H) with respect to the $z_1$ and $z_2$ redshift-samples. However, l-samples $a$ and $b$, show only small decrements, indicating that the redshift 0.3 represents an important epoch in the evolution of galaxies. As argued by Kewley et al. (2008), using a single metallicity calibration, the difference in relative metallicities should be the same using any other metallicity calibration, although the absolute metallicities might differ from one calibration to another. Thus, our main result is the relative decrement in metallicity of $\sim$0.1 dex of $z_3$ with respect to the $z_1$ and $z_2$ redshift-samples of l-sample $c$.

In order to study the behaviour of the metallicities for the different l-samples, we proceed to generate a metallicity histogram for our three l-samples as shown in Fig. 3. In the histograms we can observe a shift to lower metallicities as redshift increases, which is more evident in l-sample $c$, as argued in L09. L-samples $a$ and $b$ do not show any important variation in metallicity, as can be seen in Table 3. As a measure of the dispersion of the histograms, we estimated the interval that encompasses 66$\%$ of the galaxies around the mode of the distribution. With this criterium, we find for l-sample $a$, an interval of 0.16 dex in 12+log(O/H) for $z_0$, and  0.17 dex for $z_1$; for l-sample $b$, 0.17 dex for $z_1$, and 0.16 for $z_2$, and for l-sample $c$, 0.26 dex for $z_1$, 0.18 for $z_2$, and 0.35 for $z_3$. As can be observed in Table 3, the dispersion of the metallicity histograms increases for the samples with the more limited number of galaxies, as can be seen in l-sample $c$ for $z_1$ and $z_3$.

For the redshift--samples we can observe a maximum metallicity for $z_0$, followed by a small decrement in metallicity for redshift $z_1$ which remains constant for $z_2$, and then an important decrement for $z_3$. As indicated by Carollo et al. (2001), the redshift interval $0.5< z < 1$ represents a transition between the high-redshift universe at z $>$ 1 and that seen today. Then evolutionary effects should be more evident in the galaxy population at these redshifts. Nevertheless, in this work we find a significant metallicity evolution at redshift 0.3.

\begin{table*}[t]
\begin{center}
\begin{tabular}{cccccccccc}   
\hline
\hline
{}&\multicolumn{3}{c} {\bf l-sample $a$}&\multicolumn{3}{c} {\bf l-sample $b$}&\multicolumn{3}{c}{\bf l-sample $c$}\\\cline{5-7}
\hline
Redshift&Mean&Mode&66$\%$-range&Mean&Mode&66$\%$-range&Mean&Mode&66$\%$-range\\\hline
$z_0$&9.02&9.04&8.94-9.11&&&\\\
$z_1$&8.99&9.04&8.95-9.12&9.00&9.03&8.92-9.10&8.97&9.03&8.90-9.17\\\
$z_2$&&&&8.99&9.01&8.92-9.08&8.97&9.03&8.94-9.12\\\
$z_3$&&&&&&&8.92&8.92&8.64-8.99\\\hline
\noalign{\smallskip}
\end{tabular}
\normalsize
\rm
\end{center}
\caption{Mean and mode of the metallicity distributions, and  the metallicity range that include the 66$\%$ of the total metallicity distribution around the mode for the different luminosity samples.}
\vspace{0.1cm}
\end{table*}

\section{Nitrogen and Oxygen abundances}

The primary and secondary origin of nitrogen is of high importance to understand the processes inside stars, and the evolution of galaxies. Although this is not the aim of this paper, we are in a position to investigate the origin of nitrogen in our galaxy samples.

The nuclear mecanism producing nitrogen in stellar interiors result from the CN cycle of the CNO reactions, which takes place in the stellar hydrogen burning layer, with the net result that $^{14}$N is synthesized from $^{12}$C and $^{16}$O (Meynet $\&$ Maeder 2002; Pettini et al. 2008). Nitrogen can be of either primary or secondary origin. If the oxygen and carbon are produced in the star prior to the CNO cycling, then the amount of nitrogen produced is said to be primary. If initial amounts of oxygen and carbon are incorporated into a star at its formation, and a constant mass fraction is processed, then the amount of nitrogen produced is proportional to the initial heavy-elements abundace, and the nitrogen is said to be of secondary origin (Vila-Costas $\&$ Edmunds 1993).

Several autors (Edmunds $\&$ Pagel 1978; Barbuy 1983; Tomkin $\&$ Lambert 1984; Matteucci 1986; Carbon et al. 1987; Henry et al. 2000) demostrated that the ratio of nitrogen to oxygen remains constant at lower metallicities, $Z < 0.5$ $Z_{\odot}$ [log(O/H) + 12 $\lesssim$ 8.3, adopting 12+log(O/H)$_{\odot}$=8.66 from Asplund et al. (2005)], with a plateau at log(N/O) $\sim$ $-$1.5 in the early evolution of the galaxy, thus implying a primary origin of nitrogen. When the oxygen abundance is greater than $Z$ $\sim$ 0.5 $Z_{\odot}$, the N/O ratio rises steeply with increasing  O/H. This is the regime where nitrogen is predominantly secondary (Alloin et al. 1979; Consid\`ere et al. 2000; Pettini et al. 2008). The fact that the N/O {\bf ratio} is relatively flat at low metallicities, indicates that production of nitrogen is dominated by primary processes at low metallicities, and by secondary processes at high metallicities (Garnett et al. 1997; Ferguson et al. 1998; Henry $\&$ Worthey 1999).

In order to determine the origin of nitrogen in our l-samples, we estimated the nitrogen abundances for these galaxies. To estimate the electron temperature in the [{N\,\textsc{ii}}] emission region ($T_{[{\rm{{N_{II}}}}]}$) from log $R_{23}$, we used the formula given by Thurston et al. (1996),

 \begin{center}
$T_{[{\rm{{N_{II}}}}]}$ = 6065 + 1600(log $R_{23}$) + 1878(log $R_{23}$) + 2803(log $R_{23}$),
\end{center} with $T_{[{\rm{{N_{II}}}}]}$ in units of K. The ionic abundance ratio is estimated from the $T_{[{\rm{{N_{II}}}}]}$ temperature and the emission--line ratio ([{N\,\textsc{ii}}] $\lambda\lambda$6548, 6583)/([{O\,\textsc{ii}}] $\lambda$3727) by assuming N/O = $\rm{N}^{+}/\rm{O}^{+}$, and the flux of [{N\,\textsc{ii}}] $\lambda$6548 = (1/3) [{N\,\textsc{ii}}] $\lambda$6583. With these, we apply the formula given by Pagel et al. (1992), based upon a five--level atom calculation:

 \begin{center}
log$\left({\frac{\rm{N}^{+}}{\rm{O}^{+}}}\right)$ = log$\left({\frac{[{\rm{N}_{II}}] \lambda\lambda 6548,  6583}{[{\rm{O}_{II}}] \lambda 3727}}\right)$ + 0.307 - 0.02 log $t_{[{\rm{N}_{II}}]}$ - ${\frac{0.726}{t_{[{\rm{N}_{II}}]}}}$,
\end{center} where $t_{[{\rm{{N_{II}}}}]}$ = $10^{-4}$$T_{[{\rm{{N_{II}}}}]}$.

 \begin{figure*}[t]
 \centering
 \includegraphics[scale=0.7]{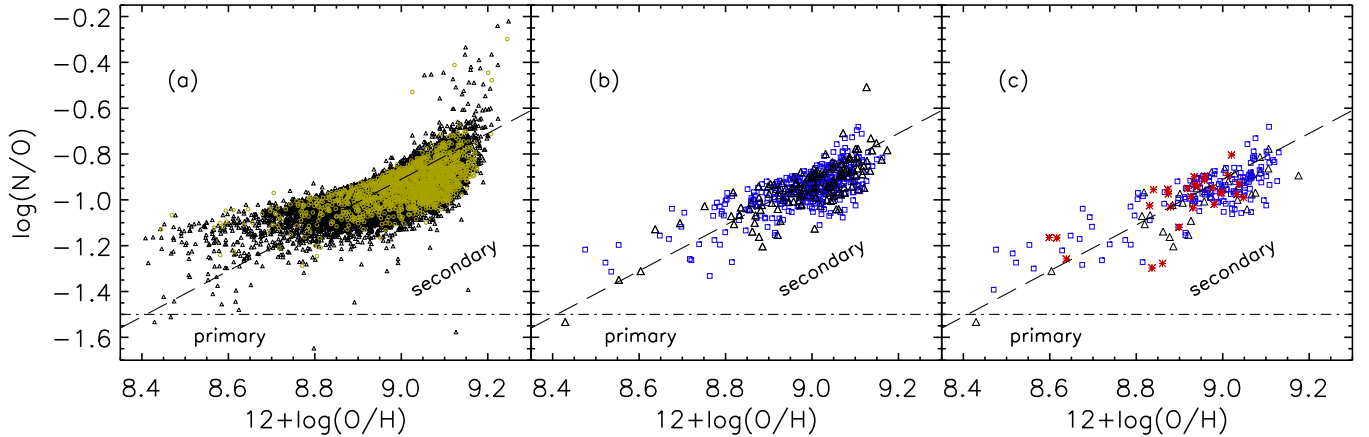}
\caption{Abundances of N and O ratios as function of their 12+log(O/H) abundances derived from the $R_{23}$ calibration for our three l-samples in redshift, following the same code of symbols used in Fig. 2. The dot--dashed line and the dashed line represent an aproximation of empirical limits of the primary and secondary levels, respectively, of N production, taken from Vila--Costas $\&$ Edmunds (1993)}
 \end{figure*}

In Fig. 4, the abundances of N and O are shown as a function of 12+log(O/H). Because nitrogen is predominantly a secondary element above metallicities about half solar, the galaxies of our samples have mainly nitrogen of secondary origin, as can be seen in Fig. 4. This is because we are working with both massive, log(M$_{star}$/M$_{\odot}) \gtrsim 10.5$ (see Fig. 6), and high metallicity galaxies. In spite of our high metallicities, we can observe signs of a horizontal population in l-sample $a$ at log(N/O) $\sim$ $-$1.2, in a range of metallicities from 8.4 to $\sim$8.6, which is a little higher than the standard value [log(N/O) $\sim$ -1.5], but comparable to the results obtained by Pettini et al. (2008) and Consid\`ere et al. (2002), who shown a population of galaxies in the same region. This result is not surprising, because, as argued above, the transition zone from primary to secondary production of nitrogen occurs at 12+log(O/H) $\sim$ 8.3, which is close to our lower limit.

Because low metallicity galaxies (12+log(O/H) $<$ 8.3) are absent in our samples, nothing can be concluded for the primary production of nitrogen.

\section{Discussion}

To investigate whether the origin of the decrement in metallicity is due to instrumental effects, to an inherent property of the sample of galaxies, or a mixture of both, it is necessary to explore two important effects: the luminosity and mass--metallicity relations, and the effect of the 3 arcsec diameter of the Sloan fibers.

\subsection{Effect of the Mass and Luminosity--Metallicity relations}

The luminosity--metallcity ($L-Z$) relation was first observed by McClure $\&$ van den Bergh (1968) in elliptical galaxies, and confirmed by  Garnett $\&$ Shields (1987), while the mass--metallicity relation was first identified for irregular and blue compact galaxies  by Lequeux et al. (1979), and Kinman $\&$ Davidson (1981), respectively, and confirmed by Skillman et al. (1989). Since then, as luminosities are easier to estimate than masses, many studies have focussed on the $L-Z$ relation (e.g., Skillman et al. 1989; Brodie $\&$ Huchra 1991; Zaritsky et al. 1994; Garnett et al. 1997; Lamareille et al. 2004, 2006; Maier et al. 2004), which correlates the absolute magnitude of galaxies with metallicity, being the more metal rich, the more luminous. The $M-Z$ and $L-Z$ relations have been studied at both, low and high redshift (e.g. T04; Savaglio et al. 2005; Erb et al. 2006; Maiolino et al. 2008; Lamareille et al. 2009).

There are two main ways to explain the origin of the $M-Z$ relation, one is attributed to metal and baryon loss due to gas outflow, where low--mass galaxies eject large amounts of metal--enriched gas by supernovae winds before high metallicities are reached, while massive galaxies have deeper gravitational potentials which helps retaining their gas, thus reaching higher metallicities (Larson 1974; Dekel $\&$ Silk 1986; MacLow $\&$ Ferrara 1999; Maier et. al. 2004; T04; De Lucia et al. 2004; Kobayashi et al. 2007; Finlantor $\&$ Dave 2008). A second scenario to explain the $M-Z$ relation is by assuming low star formation efficiencies in low--mass galaxies (Efstathiou 2000; Brooks et al. 2007; Mouhcine et al. 2008; Tassis et al. 2008; Scannapieco et al. 2008).

As pointed out in the high--resolution simulations of Brooks et al. (2007), supernovae feedback plays a crucial role in lowering the star formation efficiency in low--mass galaxies. Without energy injection from supernovae to regulate the star formation, gas that remains in galaxies rapidly cools, forms stars, and increases their metallicity too early, producing a $M-Z$ relation too flat compared to observations. However, Calura et al. (2009) reproduced the $M-Z$ relation with chemical evolution models for ellipticals, spirals and irregular galaxies, by means of an increasing efficiency of star formation with mass in galaxies of all morphological types, without the need of outflows favoring the loss of metals in the less massive galaxies. A recent study that supports this result for massive galaxies, is the one of Vale Asari et al. (2009), modelling the time evolution of stellar metallicity using a closed-box chemical evolution model. They suggest that the $M-Z$ relation for galaxies in the mass range from $10^{9.8}$ to $10^{11.65}$ M$_{\odot}$ is mainly driven by the star formation history and not by inflows or outflows.

As explained in Sec. 2, we selected small intervals of luminosity for all redshift-samples. Such selection was aimed to avoid possible biases since the SLOAN star-forming sample becomes incomplete at redshifts above z $>$ 0.1 (e.g., Kewley et al. 2006). Then, this could introduce a bias, since our high--redshift sample is formed by the most luminous galaxies, thus resulting in higher metallicity estimates.

As can be seen from Fig. 5, the galaxies of our samples do not show a luminosity--metallicity dependence, since this relation can be clearly seen only when spanning ranges of  more than $\sim$4 magnitudes in luminosity (e.g. T04). The masses of our galaxies were estimated in order to check their behaviour in our luminosity intervals with the STARLIGHT code, using a Chabrier (2003) initial mass function between 0.1 and 100 M$_{\odot}$, for details on the mass estimates, see Mateus et al (2006). The $M-Z$ diagram for our l-samples do not show any correlation either as can be seen in Fig. 6, since masses again correspond to a small interval in luminosity for each redshift bin. 

For each l-sample, we overplot the polynomial fit of the local $M-Z$ relation of Tremonti et al. (2004), which is valid over the range 8.5 $<$ log(M$_{star}$/M$_{\odot}$) $<$ 11.5, with a steep $M-Z$ relation for masses from $10^{8.5}$ to $10^{10.5}$ M$_{\odot}$, that flattens at higher masses. As can be seen in Fig. 6, galaxies of the l-sample $a$, $b$ and $c$, correspond to this flat and massive zone [log(M$_{star}$/M$_{\odot}$) $> 10.5$]  of the $M-Z$ relation, for the reasons discussed above. The mass ranges were we can found two thirds ($\sim$66$\%$) of each l-sample are shown in Table 4, as well as the median Petrosian r magnitude, and the median mass in log(M$_{star}$/M$_{\odot}$).

\begin{table*}[t]
\begin{center}
\begin{tabular}{cccccccccc}   
\hline
\hline
{}&\multicolumn{3}{c} {\bf l-sample $a$}&\multicolumn{3}{c} {\bf l-sample $b$}&\multicolumn{3}{c}{\bf l-sample $c$}\\\cline{5-7}
\hline
Redshift&$\rm{M}_r$&log(M$_{star}$/M$_{\odot}$)&66$\%$-range&$\rm{M}_r$&log(M$_{star}$/M$_{\odot}$)&66$\%$-range&$\rm{M}_r$&log(M$_{star}$/M$_{\odot}$)&66$\%$-range\\\hline
$z_0$&-21.99&11.06&10.89--11.20&&&\\\
$z_1$&-22.09&10.97&10.74--11.20&-22.99&11.44&11.28--11.58&-23.22&11.50&11.20--11.74\\\
$z_2$&&&&-23.06&11.50&11.27--11.68&-23.23&11.60&11.37--11.77\\\
$z_3$&&&&&&&-23.50&11.70&11.35--12.09\\\hline
\noalign{\smallskip}
\end{tabular}
\normalsize
\rm
\end{center}
\caption{Median Petrosian r magnitude ($\rm{M}_r$), median of the logarithm-mass [log(M$_{star}$/M$_{\odot}$)], and the logarithm-mass range that include the 66$\%$ of the total mass distribution around the median  logarithm-mass value for l-samples $a$, $b$, and $c$.}
\vspace{0.1cm}
\end{table*}

\begin{figure*}[t!]
\centering
\includegraphics[scale=0.7]{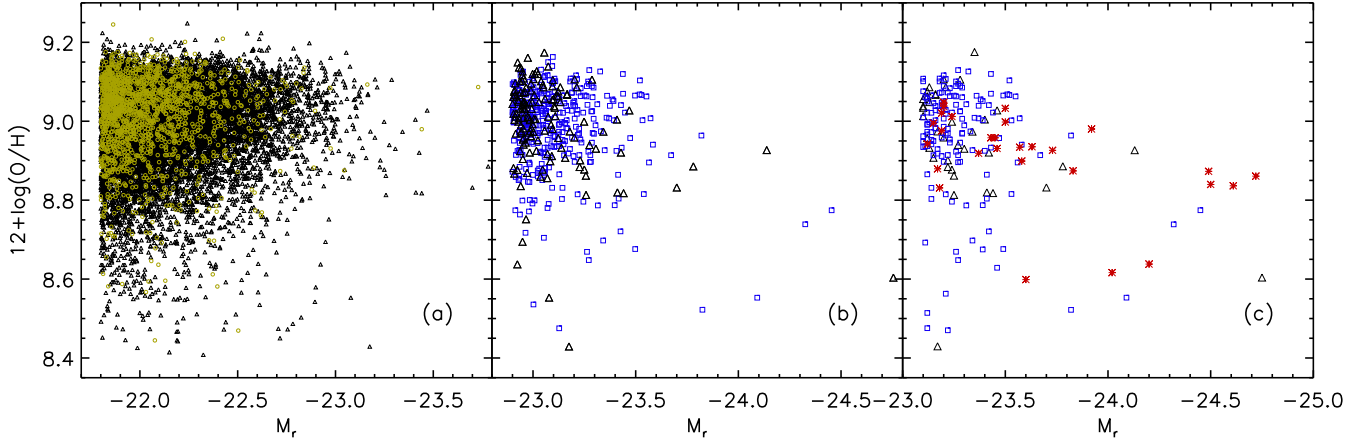}
\caption{Absolute Petrosian r magnitude and 12+log(O/H) for our l-sample of galaxies ($L-Z$ relation). Symbols follow the same code used in Fig. 2.}
\end{figure*}

\begin{figure*}[t!]
\centering
\includegraphics[scale=0.7]{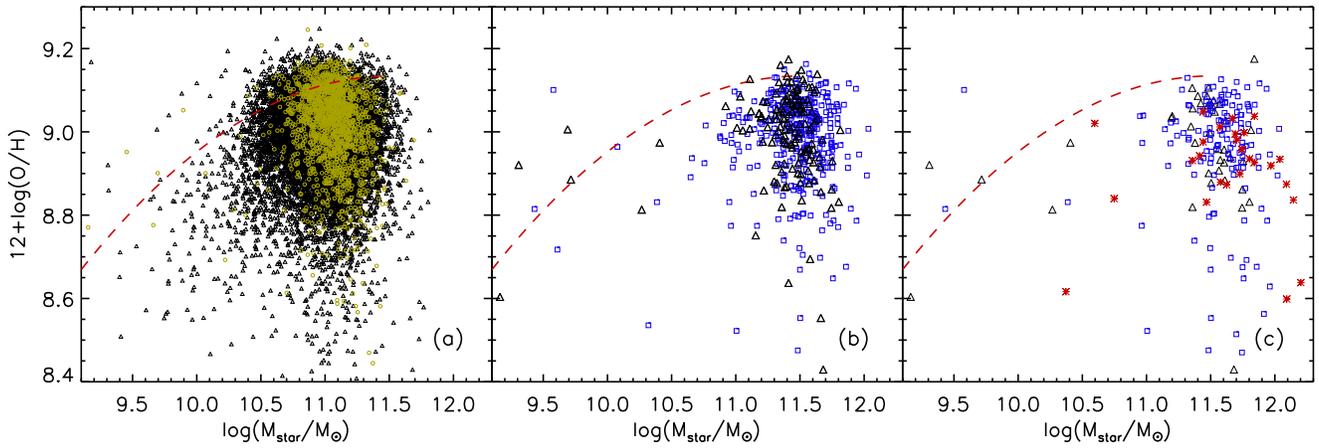}
\caption{Relation between stellar mass, in units of solar masses, and 12+log(O/H) for our l-sample of galaxies ($M-Z$ relation). Dashed line represent the fit of Tremonti et al. (2004), and symbols follow the same code used in Fig. 2.}
\end{figure*}

\subsection{Effects of the Sloan fiber diameter}

With respect to the Sloan fiber diameter, and depending on the galaxy size, we expect that at higher redshift the projected size of the Sloan fibers will cover a larger fraction of galaxy area than for nearby galaxies. This effect, as argued in Sect. 3, could introduce a bias in our samples since integrated metallicities are lower than nuclear ones. In order to quantify this contribution, we estimate the percentage of angular size of each galaxy inside the three arcsec diameter of the Sloan fiber using the Petrosian total radius in r--band in arcsec. To this aim, we divide the fibre radius (1.5 arcsec) by the Petrosian total radius in r--band (petrorad$_r$), as can be seen in the histograms of Fig. 7 for all our l-samples. Thus, this ratio can be taken as the fraction of the galaxy size that is actually covered by the fibre.

As expected, in all l-samples the distribution shifts to a maximum coverage of the galaxy size as redshift increases. This is an effect that must be taken into account, because it could change the metallicity estimation according to the fraction of galaxy diameter inside the Sloan fiber. As argued in L09, the decrement observed in metallicity for l-sample $c$  can not be atributed to the percentage of galaxy inside the Sloan fiber, because fibers cover less than 50$\%$ of galaxy sizes for $\sim$95$\%$ of this sample. However, the effect could be noticeable at lower redshifts.

\begin{figure*}[t!]
\centering
\includegraphics[scale=0.5]{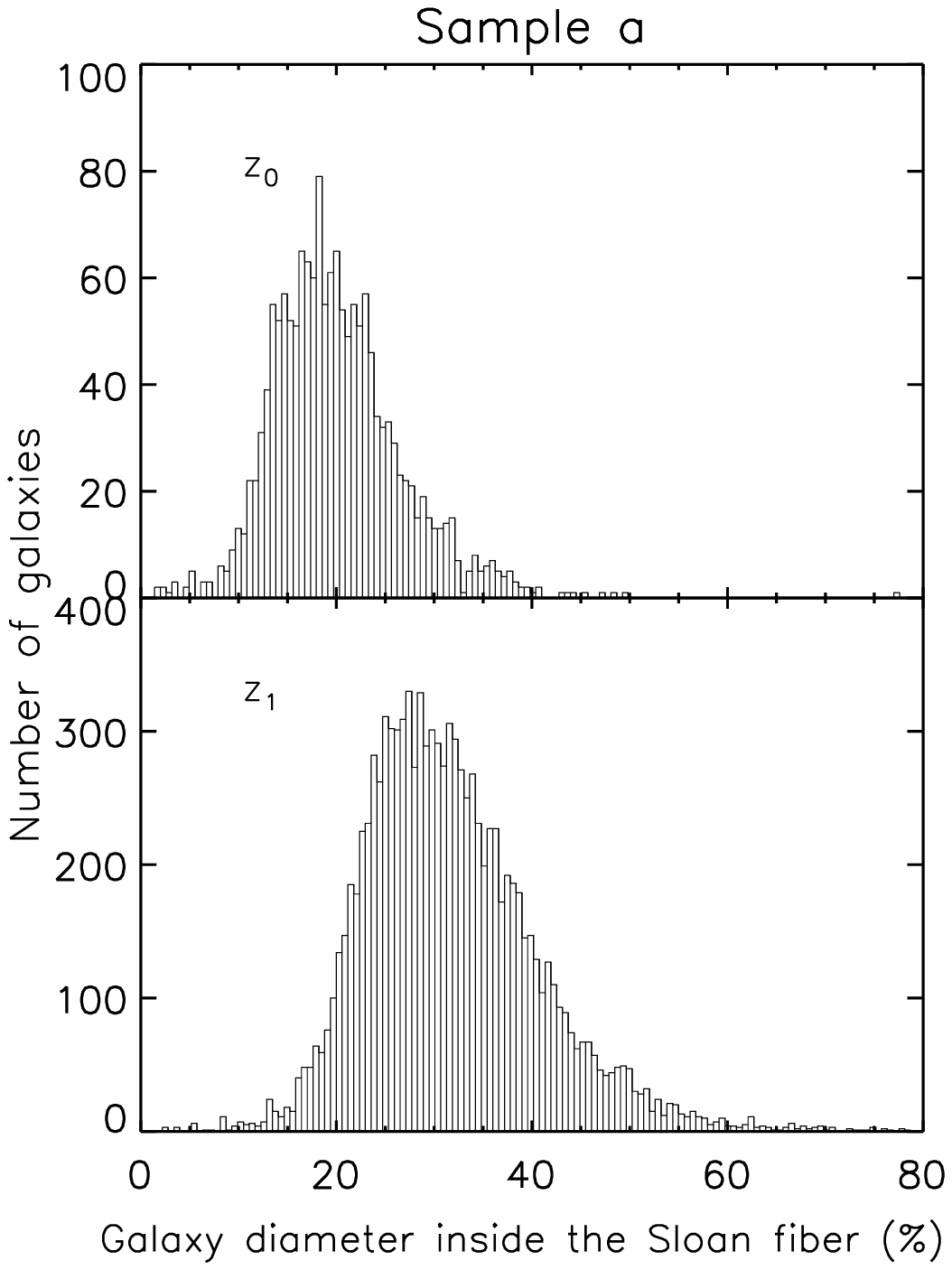}
\includegraphics[scale=0.5]{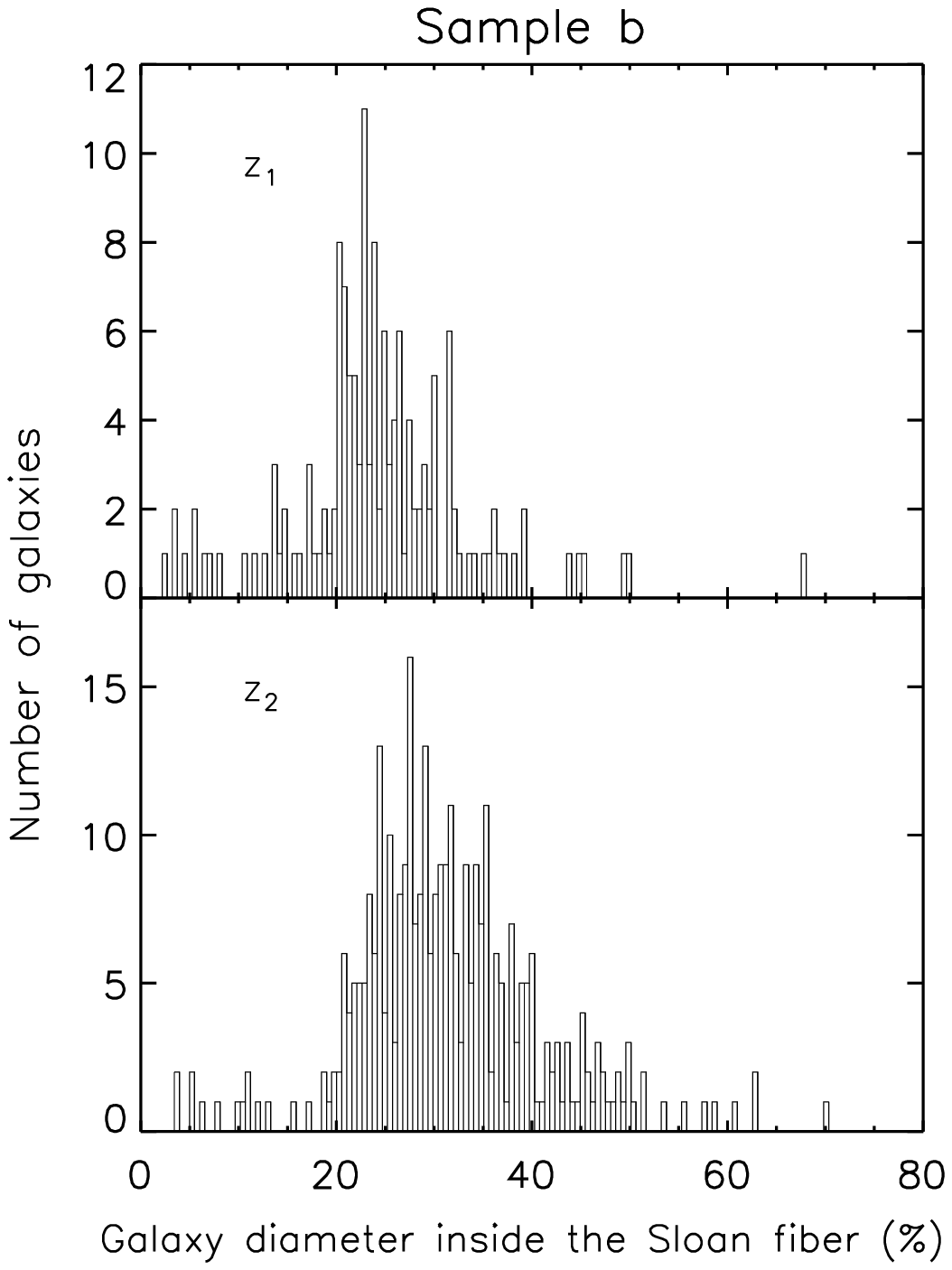}
\includegraphics[scale=0.5]{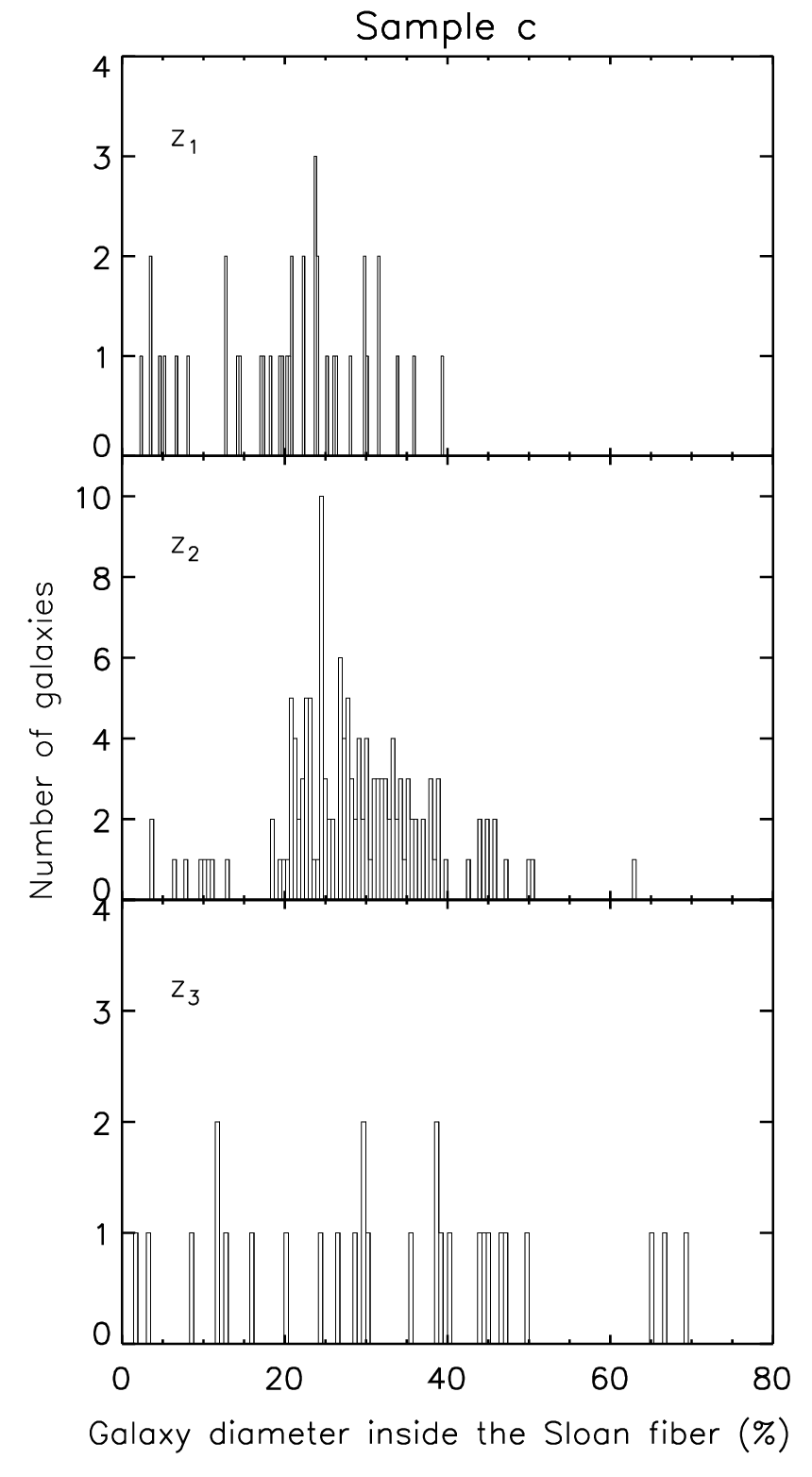}
\caption{Percentage of angular size of each galaxy inside the three arcsec diameter of the Sloan fiber for l-samples $a$, $b$, and $c$.}
\end{figure*}

To minimize systematic errors from this aperture biasses, as explained in Sec. 2, we selected galaxies with  $z > 0.04$ following the recommendations of Kewley et al. (2005), who investigate the effects of a fixed--size aperture on spectral properties for a large range of galaxy types and luminosities, concluding that a minimum flux covering fraction of 20$\%$, corresponding to a median redshift of z $\sim$ 0.04, is required for metallicities to approximate the global values.

It is important to point out that in Fig. 7 we are plotting the percentage of angular size of each galaxy inside the three arcsec diameter of the Sloan fiber, which is not the same as the flux covering fraction used by Kewley et al. (2005).

However, in spite of this redshift limit and in order to test how much the percentage of angular size inside the Sloan fiber diameter affect our metallicities, we compare in Fig. 8 our original metallicity estimates of redshift--sample $z_0$, with that from galaxies with a percentage of angular size within the fiber $>$20$\%$, which corresponds to a 45$\%$ of the sample. We observe that the result is the same, with a quite small difference in the mean of the order of $\sim$0.001 in 12+log(O/H)  for galaxies with a percentage of angular size within the fiber $>$20$\%$.

If redshift sample $z_0$, which contains the maximum fraction of galaxies with a percentage of angular size  $<$20$\%$ inside the Sloan fiber diameter, do not show variations in the mean metallicity, we can assume that none of our samples are significantly affected by aperture effects due to the Sloan fiber diameter.

\begin{figure}[t!]
\centering
\includegraphics[scale=0.5]{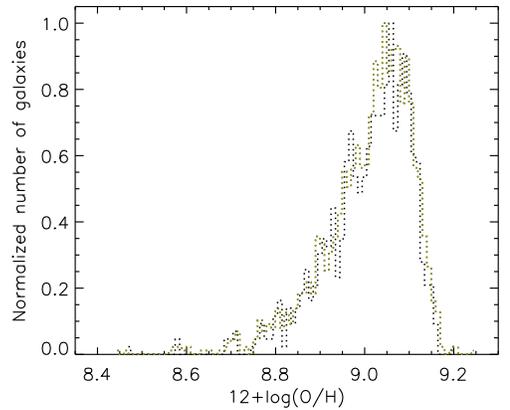}
\caption{Normalized metallicity histograms for redshift-sample $z_0$. Dotted clear line is the same as in Fig. 3a, and dotted dark line are galaxies of the same redshift-sample with covering fraction $>$20.}
\end{figure}


\section{Conclusions}

Although similar studies of metallicity using SDSS exists, they are either restricted to a redshift $\sim$0.1 (e.g. T04), or do not segregate their samples as a function of redshift (e.g. L06), thus making not possible to detect metallicity evolution at low redshift. On the other hand, studies at high redshift are statistically limited. 

We divided our sample in redshift intervals of $\Delta$z $\sim$0.1 with the goal of identifying any evolution in metallicity, and each redshift in small intervals of luminosity in order to avoid biases due to the luminosity and mass-metallicity relations. Because we are comparing galaxies from redshift 0.04 to 0.4, we selected luminous and massive galaxies [$\sim$ log(M$_{star}$/M$_{\odot}$) $> 10.5$], which are present in all redshift intervals.

We conclude that the nitrogen production for our sample of galaxies is mainly secondary, because our sample is formed by massive, luminous and high metallicity galaxies. We remark that in this work we can not conclude nothing about the production of nitrogen in low massive galaxies.

We derived the $M-Z$ and $L-Z$ relations for our sample of galaxies, showing a flat $L-Z$ relation, and a $M-Z$ relation populated only in the massive zone $\sim$ log(M$_{star}$/M$_{\odot}$) $> 10.5$, which is due to the absence of the dwarf galaxies. Both relations behave as expected since our sample is selected to cover the more luminous and hence massive galaxies.

Since we have metallicity estimates for redshifts up to 0.4 in bins of 0.1, we are able to represent redshift versus metallicity, as shown in Fig. 9. We plotted the mean metallicity for all l-samples $a$, $b$, and $c$. Error estimates of the mean metallicity were obtained from the line error fluxes provided by STARLIGHT, taking into account the number of galaxies of each sample.

As shown in Fig. 9, we can observe, for l-sample $a$, an initial metallicity of $\sim$9.02 for $z_0$, followed by a small decrease in metallicity for $z_1$. For l-sample $b$, metallicity of $z_1$ and $z_2$ remains constant, a trend also observed for l-sample $c$ in the same redshift--samples, but with lower metallicity values, and finally an important decrement for $z_3$.  We overplot the model of Buat et al. (2008) for galaxies with a rotational velocity of 360 km/s, which correspond to a log(M$_{star}$/M$_{\odot}$) $\sim$ 11.2 in his model. 

Since the model of Buat et al. (2008) is calibrated in solar metallicities, with log(Z/Z$_{\odot}$) $\sim$ $-$0.03 at redshift zero, we assume a solar metallicity of 12+log(O/H)$=$9.05 in order to match his fit to our metallicity values, finding a good correlation between her model and our metallicity dependence versus redshift. Finally, we fit a second order polynomial (see Fig. 9).

It is important to have in mind that this result is valid for massive, luminous, and high metallicity galaxies. Lower metallicity and stellar mass galaxies are absent from our sample due to the selection criteria applied.

Although it is well known that metallicities decrease with redshift, it is the first time that a statistically significant sample of galaxies is analized looking for metallicity evolution at such a low redshift, detecting small decrements as redshift increase, with $prima facie$ evidence of a quite important decrement at the redshift--interval 0.3-0.4, as already pointed out in L09.

\begin{figure}[h]
\centering

\includegraphics[scale=0.60]{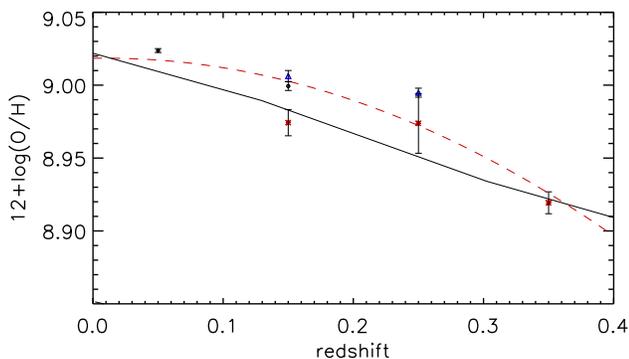}
\caption{Metallicity evolution derived from our l-samples up to redshift 0.4, vertical lines show the error with respect to the mean metallicity for every redshift interval. Circles, triangles and asterisks, represent the mean metallicity of the l-samples $a$, $b$ and $c$, with its respective redshift-samples. Solid  line represents the model of Buat et al. 2008 for galaxies with a rotational velocity of 360 km/s, and dashed line represents a second-order polynomial fit (y=$a_0+a_1x+a_2x^2$), with $a_0=9.018$, $a_1=0.015$, and $a_3=-0.799$.}
\end{figure}

\begin{acknowledgements}
This work was supported by the Spanish
\emph{Plan Nacional de Astronom\'{\i}a y Astrof\'{\i}sica} under grant AYA2008-06311-C02-01. The Sloan Digital Sky Survey (SDSS) is a joint project of The University of Chicago, Fermilab, the Institute for Advanced Study, the Japan Participation Group, The Johns Hopkins University, the Max--Planck--Institute for Astronomy, Princeton University, the United States Naval Observatory, and the University of Washington. Apache Point Observatory, site of the SDSS, is operated by the Astrophysical Research Consortium. Funding for the project has been provided by the Alfred P. Sloan Foundation, the SDSS member institutions, the National Aeronautics and Space Administration, the National Science Foundation, the U.S. Department of Energy, and Monbusho. The official SDSS web site is www.sdss.org. We thank to the STARLIGHT Project Team (UFSC, Brazil), specially to William Schoenell, who made the best efforts to help us in the download of the whole data set. We thank to Veronique Buat and Samuel Boissier for providing us the data of his metallicity evolution model. We also thank to the anonymous referee for his/her meticulous revision of the manuscript and for the constructive comments. Maritza A. Lara-L\'opez is supported by a CONACyT and SEP fellowships.

\end{acknowledgements}

\end{document}